\documentclass[12pt]{amsart}

\usepackage{epsfig}

\begin{document}

\title[]{\bf Radiative Contributions to the Effective Action of Self-Interacting Scalar Field on a 
           Manifold with Boundary}

\author[]{George Tsoupros \\
       {\em School of Mathematics and Statistics}\\
       {\em The University of Sydney}\\
       {\em NSW 2006}\\
       {\em Australia}
}
\thanks{present e-mail address: gts@newt.phys.unsw.edu.au}

\begin{abstract}

The effect of quantum corrections to a conformally invariant field theory for
a self-interacting scalar field on a curved 
manifold with boundary is considered. The analysis is most easily performed 
in a space of constant curvature the boundary of which is characterised by 
constant extrinsic curvature. 
An extension of the spherical formulation in the presence of a boundary 
is attained through use of 
the method of images. Contrary to the consolidated vanishing effect in maximally 
symmetric space-times 
the contribution of the massless `` tadpole'' diagram no longer vanishes in 
dimensional regularisation.
As a result, 
conformal invariance is broken due to boundary-related vacuum contributions. 
The evaluation of one-loop contributions to the two-point function suggests 
an extension, in the presence of matter couplings, of the simultaneous volume 
and boundary renormalisation in the effective action. 
 
\end{abstract}

\maketitle

{\bf I. Introduction}\\

The investigation of elliptic operators on Riemannian manifolds with boundary 
is a pivotal aspect in Euclidean Quantum Gravity. The eigenvalue problem for 
operators of Laplace type, in particular, arises naturally in the context of 
the semiclassical approximation to the wave function of the Universe. 
Specifically, instanton related considerations in quantum and inflationary 
cosmology lend particular importance to the issue of radiative contributions 
to a semiclassical tunneling geometry of constant curvature, effected by 
quantised matter. In the case of a quantised scalar field coupled to the 
background geometry such an issue relates naturally to the eigenvalue problem 
of the Laplace operator formulated on a bounded segment of the de Sitter 
four-dimensional sphere. For this spatially symmetric situation it is possible 
to relate the eigenvalue problem on the bounded region $ \it{C}$ of the 
spherical cap to one on its covering manifold which constitutes the entire 
four sphere. The method of images can accomplish this task provided that either 
Dirichlet or Neumann (but not Robin) conditions are specified on the boundary 
for the scalar field \cite{Dowker}. The latter provide also a sufficient 
condition for self adjointness of the Laplace operator. 

As a consequence of the investigation of elliptic operators heat kernel
and functional methods have been invoked for the evaluation of the one-loop 
semiclassical approximation to the functional integral for quantum gravity, 
that is to the formal sum over a specified set of geometries meeting the 
boundary conditions \cite{Barvinsky}, \cite{Esposito}. These methods have been extended
in the presence of matter couplings \cite{Moss}, \cite{Od}, \cite{Odin}, \cite{Odintsov}.
Although use of renormalisation group techniques have yielded the improved effective 
action past one-loop order \cite{Odintsov} no attempt 
has hitherto been made for the evaluation of higher loop-order contributions 
to semiclassical tunnelling geometries past that level. 
Such calculations would necessarily rely on diagramatic techniques on a 
manifold with boundary. The relevant approach would be predicated on the concept of the 
relativistic propagator and would be distinct from diagramatic techniques related to
heat kernel asymptotic expansions hitherto proposed \cite{BV}, \cite{MW}.
Fundamental in this calculational context is the 
evaluation of the contribution which the boundary of the manifold has to 
the propagator of the relevant quantised matter field coupled to the 
manifold's semiclassical background geometry. In the physically important 
case of a n-dimensional bounded spherical cap $ \it{C_n}$ the method of 
images can be invoked as the computational technique for the matter propagator. 
In relating the relevant eigenvalue problem on $ \it{C_n}$ to one on its 
covering manifold the method of images essentially relates propagation on 
$ \it{C_n}$ to propagation on the entire $ S_n $. This is the case because the 
propagator is necessarily the Green function associated with the bounded 
elliptic operator on $ \it{C_n}$. Specifically, in the case of massless 
propagation on $ \it{C_n}$ for a scalar field conformally coupled to the 
background geometry both the fundamental part of the Green function and the 
additional to it part arising from boundary conditions specified on 
$ \partial C_n$ admit an expansion in terms of the eigenfunctions of the 
Laplace operator defined on $ S_n$ provided that the method of images is used. 
In effect, a consistent use of the method of images dispenses with the rather 
intractable expansions involving spherical harmonics of fractional degrees 
which naturally arise as eigenfunctions of the bounded Laplace operator 
on $ \it{C_n}$. 

The present work addresses the issue which the presence of a 
$ (n-1)$-dimensional  boundary of constant extrinsic curvature on a Riemannian 
n-manifold of constant curvature raises for the dynamical behaviour of a 
conformal scalar field defined on that manifold. The presence of such 
boundaries no longer allows for the established spherical formulation of 
conformal scalar theories as it destroys the SO(n+1) (de Sitter) invariance on 
which the latter are predicated. It will be shown that in relating the 
eigenvalue problem for the bounded Laplace operator on the n-cap to that on the 
entire n-sphere the method of images allows for a simple solution to the Green 
equation with a direct physical interpretation. As a precursor to a 
renormalisation program a conformal scalar field supplemented with a 
self-interacting $ \phi^4$ coupling will be considered in the $ n=4 $ case. 
It will be shown that the vanishing effect which the technique of dimensional 
regularisation is known to have on the massless ``tadpole'' diagram on maximally 
symmetric manifolds is no longer sustainable in the present case of the 
stated bounded manifold, a result which suggests general significance for that 
vanishing effect only on maximally symmetric manifolds. The boundary effect on 
the ``tadpole'' diagram is finite as a consequence of which to first order in 
the self-coupling the scalar vacuum effects do not result in infinite contributions 
to the semi-classical boundary part of the bare action. At higher orders,
nevertheless, more complicated diagramatic structures entail the potential for 
generation of infinite redefinitions. It will be shown that the relevant loop 
integrals have the potential for the simultaneous redefinition of the 
volume-related terms in the bare action and of those terms in the latter which 
the presence of the boundary $ \partial \it{C_n}$ necessitates at the semi-classical 
level as a condition for the validity of the Einstein equations on $ \it{C_n}$. 
This allows for the generalisation to arbitrary loop-orders of the simultaneous 
renormalisation of boundary and volume ultra-violet divergences attained at one-loop 
level through use of heat kernel techniques \cite{Solodukhin}. Moreover, the volume 
contribution of the ``tadpole'' to the effective action will be shown to have the 
potential for generation of conformally non-invariant counterterms at higher orders.

{\bf II. The method of images and propagation on the bounded manifold}\\

The spherical formulation of the dynamical behaviour of a conformal scalar 
field $\phi$ admitting a classical action which remains invariant under 
the conformal rescaling 

\begin{equation}
g_{\mu\nu} \rightarrow \Omega^{2}(x)g_{\mu\nu},~~~
\phi \rightarrow \Omega^{1-\frac{1}{2}n}\phi \equiv \Phi
\end{equation}
in a general n-dimensional Lorentzian space-time, is attained on a Riemannian 
manifold by specifying its background geometry to be that of positive 
constant curvature $ r$ embedded in a $ (n+1)$-dimensional space. 
The dynamical behaviour of self-interacting scalar models has been studied 
in this context for $ n =4$ and $ n=3$ \cite{Drummond}, 
\cite{McKeon Tsoupros}, \cite{Tsoupros} by exploiting the coincidence between 
the classical action of the theory specified on the n-sphere $ S_n$ and the 
classical action obtained by conformally mapping the theory specified on the 
n-dimensional Euclidean space, onto $ S_n$. The spherical n-cap $ \it{C_n}$ 
considered as a manifold of constant curvature embedded in a 
$ (n+1)$-dimensional Euclidean space and bounded by a $ (n-1)$-sphere of 
positive extrinsic curvature $ k$ (diverging normals) is, in the same respect, 
conformal to the interior of the associated $ (n-1)$-sphere which constitutes 
the n-disk embedded in the same $ (n+1)$-space. The choice of transformation 
(1) which maps Euclidean space-time onto a de Sitter sphere and thereby the 
n-disk onto the n-cap is, for many purposes, arbitrary. The usual technique 
of stereographic projection will be invoked for the purposes of the ensuing 
analysis. In conformity with the results in \cite{Drummond}, a massless 
scalar field $ \phi $ on the n-dimensional disk with the Dirichlet condition 
$ \phi = 0 $ specified on the disk's boundary is mapped on the n-cap into the 
spherical form  

\begin{equation}
\Phi = \kappa^{1-\frac{n}{2}} \phi
\end{equation} 
with the multiplicative constant 

\begin{equation}
\frac{\partial (\eta)}{\partial (x)} = \kappa^{n+1}
\end{equation}
relating to the Jacobian of the conformal transformation 
$ x \rightarrow \eta $, $ x$ and $ \eta$ being $ n+1$ vectors. The same 
condition $ \Phi = 0$ on the cap's $ (n-1)$-dimensional spherical boundary 
emerges naturally through this transformation. In effect, the kinetic part of 
the scalar action on the cap becomes

\begin{equation}
\frac{1}{2}\int{d^n\eta}{\Phi}{M_c}{\Phi} = 
\frac{1}{2}\int{d^nx}{\phi}{\partial}^2{\phi} 
\end{equation}
where integration is understood over the cap and the disk volume respectively. 
$ M_c$ is the relevant bounded spherical Laplace operator conformally related 
to the d'Alembertian $ {\partial}^2 $, its exact form remaining the same as 
that of the unbounded Laplace operator $ M$ defined on the entire $ S_n$

\begin{equation}
M_c = D^2 - \frac{n(n-2)}{4R^2};~~~ 
D_a = (\delta_{ab} - \frac{{\eta_a}\eta_b}{R^2})\frac{\partial}{\partial{\eta_a}}
\end{equation}
but its domain being different due to the presence of the boundary. In effect, 
the spherical harmonics $ Y^N_{\alpha}$ of integral degrees of homogeneity 
$ N$ which in $ n+1$ dimensions form a complete set of eigenfunctions for the 
corresponding Laplacian defined on $ S_n$ are no longer eigenfunctions of 
$ M_c$. The non trivial boundary conditions alter the spectrum of eigenvalues 
thereby directly affecting the corresponding degrees $ N$ which become, in 
turn, fractional as a condition for orthonormality. It has been shown, in 
fact, that cap spherical harmonics do not actually form a complete set 
\cite{Haines}.

It is evident at this point that the presence of a boundary on a manifold of 
constant curvature is incompatible with a direct application of the spherical
formulation. Moreover, the stated lack of completeness and the emergence of 
fractional values for the degrees $ N$ which are physically associated with 
angular momenta flowing through the relevant propagators would tend to 
obscure a direct physical interpretation of any perturbative calculation.
Such complications are the direct result of the smaller symmetry group
which is retained for the Riemannian bounded manifold from the original 
SO(n+1) invariance. It would, for that matter, be desirable to relate the 
eigenvalue problem for the bounded Laplace operator $ M_c$ defined on 
$ \it{C_n}$ to that of the unbounded Laplacian $ M$ on $ S_n$, the covering 
manifold of $ \it{C_n}$. The method of images is the simplest expedient to 
this end and can be best applied on the embedded n-dimensional disk which is 
conformal to $ \it{C_n}$ rather than on $ \it{C_n}$ itself. Specifically, for 
$ n>2$ the Green's function to the Euclidean n-dimensional $ \partial^2$ is 
$ |x-x'|^{2-n}$. The presence of the disk's boundary on which the condition 
$ \phi = 0$ has been specified generates an additional term which remains 
finite at the limit $ x \rightarrow x'$, for any $ x'$ in the volume of the 
disk. In effect, the massless scalar propagator on the n-disk is

\begin{equation}
D^{(n)}(x,x') = \frac{1}{|x-x'|^{n-2}} - 
\frac{1}{|\frac{|x'|}{r}x-\frac{r}{|x'|}x'|^{n-2}}
\end{equation}
with $ r$ being the radius of the $ (n-1)$-sphere which constitutes the 
disk's boundary. This propagator is in conformity with the stated Dirichlet 
condition and reduces to the flat-space propagator for the massless scalar 
field at the $ r \rightarrow \infty$ limit. 

The propagator on $ \it{C_n}$ can now be obtained by exploiting the conformal 
relation between the latter and the n-disk. Crucial to the general 
significance of this approach is the aforementioned equivalence between the 
action for the scalar field conformally coupled to the background geometry of 
$ \it{C_n}$ and that obtained through a conformal transformation of the 
action for a massless scalar field specified on the n-disk 
\cite{McKeon Tsoupros}. The stereographic projection applied with reference to
the north pole on $ S_n$ maps the centre of the n-disk onto the north pole's 
counterdiametric pole, that is, on the pole of $ \it{C_n}$. In effect, $ r$ is
mapped onto the geodesic distance $ a_B$ between the cap's pole and any point 
on the boundary $ S_{n-1}$ of the cap. As a result, $ D^{(n)}(x,x')$ is mapped 
onto 
 
\begin{equation}
D_{c}^{(n)}(\eta,{\eta}') = \frac{1}{|{\eta}-{\eta}'|^{n-2}} - 
\frac{1}{|\frac{a_{{\eta}'}}{a_B}{\eta}-
\frac{a_B}{a_{{\eta}'}}{\eta}'|^{n-2}}
\end{equation}
with $ a_{{\eta}'}$ being the geodesic distance between the cap's pole
and $ {\eta}'$. $ D_{c}^{(n)}$ is the desired propagator for a conformal 
scalar field on $ \it{C_n}$. It possesses the same structure as that 
of its conformal counterpart and satisfies 

\begin{equation}
{M_c}D_{c}^{(n)}(\eta,{\eta}') = {\delta}^{(n)}({\eta}-{\eta}')
\end{equation}
with the condition $ D_{c}^{(n)}(\eta,\eta')=0$ when $ a_{\eta'}=a_{B}$ 
which guarrantees the absence of propagation on the boundary hypersurface. 
The fundamental part $ |{\eta}-{\eta}'|^{n-2}$ of this Green function is the 
usual Euclidean de Sitter space propagator $ D_{s}^{(n)}(\eta,{\eta}')=0$. 
The additional to it boundary term expresses the contribution due to 
reflection off the boundary hypersurface and remains finite at the coincidence 
limit $ \eta \rightarrow {\eta}'$ in the volume of $ \it{C_n}$. It can be seen, 
however, to develop the same divergence at the coincidence limit on 
$ \partial C_n$ as that of the fundamental part. This divergence is a direct 
consequence of the demand for vanishing propagation on the boundary and will be 
shown to be responsible for the presence of $ \Phi$-related boundary terms in 
the effective action in spite of the Dirichlet condition $ \Phi=0$ on 
$ \partial C_n$. It is also worth noting that $ D_{c}^{(n)}$ reduces to 
$ D_{s}^{(n)}$ for $ a_{{\eta}'}=0$, that is, for any propagation to the pole 
of $ \it{C_n}$. The physical significance of this effect, which is also 
characteristic of massless scalar propagation on the n-disk, stems from the 
invariance group of $ \it{C_n}$ and apparently refers to the unique cancellation 
of all contributions from the boundary to pole-directed propagation. 

The fundamental part of $ D_{c}^{(n)}(\eta,{\eta}')$ is the elementary Haddamard
function admitting the expansion

\begin{equation} 
|{\eta}-{\eta}'|^{2-n} = \sum_{N=0}^{\infty}\sum_{\alpha=0}^{N}
\frac{1}{\lambda_N}Y_{\alpha}^N({\eta})Y_{\alpha}^N({\eta}')
\end{equation}
in terms of the complete set of spherical harmonics in $ n+1$ dimensions
\cite{Drummond}

\begin{equation}
\int{d^n\eta}Y_{\alpha}^N({\eta})Y_{{\alpha}'}^{N'}({\eta}) = 
\delta_{NN'}^{(n)}\delta_{\alpha {\alpha}'}^{(n)}
\end{equation}

\begin{equation}
\sum_{N=0}^{\infty}\sum_{\alpha=0}^{N}
Y_{\alpha}^N({\eta})Y_{\alpha}^N({\eta}') = \delta^{(n)}({\eta}-{\eta}')
\end{equation}
which are eigenfunctions of the Laplacian $ M$ on the embedded $ S_n$ of radius $ a$, 
the covering manifold of $ \it{C_n}$ 

\begin{equation}
MY_{\alpha}^N({\eta}) = \lambda_NY_{\alpha}^N({\eta})
\end{equation}

\begin{equation}
\lambda_N = -\frac{(N+\frac{n}{2}-1)(N+\frac{n}{2})}{a^2}
\end{equation}

At the same time, the expression $ |\frac{a_{{\eta}'}}{a_B}{\eta}-
\frac{a_B}{a_{{\eta}'}}{\eta}'|$ in the boundary part of the propagator in (7)
is the geodesic distance between the associated two points 
$ \frac{a_{{\eta}'}}{a_B}{\eta}$ and $ \frac{a_B}{a_{{\eta}'}}{\eta}'$. In 
fact, $ \frac{a_B}{a_{{\eta}'}}>1$ indicates that this geodesic distance is 
relevant to the entire $ S_n$ rather than merely to $ \it{C_n}$. In conformity 
with the method of images this allows for the interpretation of the boundary 
contributions to massless scalar propagation between $ \eta$ and 
$ {\eta}'$ on $ \it{C_n}$ as arising from propagation on $ S_n$. However, the 
condition of a lower, non-vanishing limit is enforced on this propagation by the 
coincidence limit $ \eta \rightarrow \eta'$ on $ \it{C_n}$. The propagation on 
$ S_n$ representing the boundary contributions does not occur for geodesic 
separations smaller than 
$ |\frac{a_{{\eta}'}}{a_B}{\eta}'- \frac{a_B}{a_{{\eta}'}}{\eta}'|$. In effect, 
the boundary part in (7) being itself an elementary Haddamard function admits the 
same expansion

\begin{equation} 
|\frac{a_{{\eta}'}}{a_B}{\eta}-
\frac{a_B}{a_{{\eta}'}}{\eta}'|^{2-n} = 
\sum_{N=0}^{N_0}\sum_{\alpha=0}^{N}
\frac{1}{\lambda_N}Y_{\alpha}^N(\frac{a_{{\eta}'}}{a_B}{\eta})
Y_{\alpha}^N(\frac{a_B}{a_{{\eta}'}}{\eta}')
\end{equation}
with the condition that now only a finite number of terms in the complete set 
${Y_{\alpha}^N({\eta})}$ is relevant as a result of the cut-off separation. The 
integer degree $ N$ in (14) is physically associated in transform space with 
the quantum number for the angular momentum flowing through the image-propagator 
on $ S_n$ which the boundary part signifies in the same way as $ N$ in (9) is 
physically associated with the quantum number for the angular momentum flowing 
through the propagator $ D_{s}^n(\eta,{\eta}')$ on $ \it{C_n}$ which the 
fundamental part of $ D_{c}^{(n)}(\eta,{\eta}')$ signifies. The cut-off scale 
in transform space which $ N_0$ signifies is evidently an increasing function 
of the geodesic distance $ a_B$.

The expansions (9) and (14) make it evident that the reduction of the 
eigenvalue problem on $ \it{C_n}$ to an eigenvalue problem on $ S_n$ through 
the method of images allows for the exploitation of the spherical formulation 
in the context of any perturbative calculation on $ \it{C_n}$ although the 
propagator $ D_{c}^{(n)}(\eta,{\eta}')$ on the latter no longer admits a direct 
expansion of the form (9). In order to exemplify the merit of this approach 
as a preamble to perturbative renormalisation on $ \it{C_4}$ the ``tadpole'' 
diagram will be evaluated in what follows.

{\bf III. The massless tadpole and dimensional renormalisation}\\

The simplest diagrams in the perturbative expansion of the effective action 
are those representing contributions which depend on $ D_{c}^{(n)}(0)$ 
(``bubbles'' and ``tadpoles''). They are known to vanish in all maximally 
symmetric space-times provided that dimensional regularisation is used 
\cite{Drummond}. That regulating technique manifests all divergences as 
poles at the limit of the relevant space-time dimensionality $ n$ through 
an analytical extension of the latter in the complex plane. The vanishing 
effect in its context is due to a peculiar cancellation between the 
ultra-violet and the infra-red divergence in any massless scalar theory. In 
the present case, the method of images allows for the treatment of both 
fundamental and boundary part in $ D_{c}^{(n)}(\eta,{\eta}')$ as propagation 
on the entire $ S_n$ thereby calling naturally into question the persistence 
of the stated vanishing effect on $ \it{C_n}$. 

The loop integral in configuration space for any diagram representing 
$ D_{c}^{(n)}(0)$-related contributions is

\begin{equation}
I(n) = \int_{C}{d^n}{\eta}D_{c}^{(n)}(\eta,{\eta})
\end{equation}
with integration over $ \it{C_n}$. Its fundamental-part related term 

\begin{equation}
\int_{C}{d^n}{\eta}\frac{1}{|{\eta}-{\eta}'|^{n-2}}
\end{equation}
is formally divergent at $ n=4$ when the $ \eta \rightarrow \eta'$ limit is 
considered in the integrand. The evaluation of this term, however, necessitates 
$ n<2$ from the outset. A subsequent analytical extension causes this integral 
to vanish when continued back to $ n=4$. Nevertheless, the boundary-related term 
of the same loop integral

\begin{equation}
\int_{C}{d^n}{\eta}\frac{1}{|\frac{a_{{\eta}'}}{a_B}{\eta}- 
\frac{a_B}{a_{{\eta}'}}{\eta}'|^{n-2}}
\end{equation}
is finite and non-vanishing at $ \eta \rightarrow \eta'$ when continued back to 
$n=4$. As a result, all $ D_{c}^{(n)}(0)$-related contributions to the effective 
action, are no longer vanishing. The underlying source of these contributions 
can be traced to the cut-off separation characterising the boundary part of 
$ D_{c}^n(\eta,{\eta}')$ in (17). Specifically, the following relation between 
spherical harmonics defined on a Euclidean n-sphere of radius $ a$ and Gegenbauer 
polynomials 

\begin{equation}
\sum_{\alpha=0}^{N}Y_{\alpha}^N({\eta})Y_{\alpha}^N({\eta}') = 
F(N,n)C_{N}^{\frac{n-1}{2}}(z)
\end{equation}
with 

\begin{equation}
z=\frac{{\eta}.{\eta}'}{a^2} ;~~~F(N,n)=\frac{2N+n-1}{4a^n \pi^{\frac{n}{2}}}
\Gamma(\frac{n-1}{2})
\end{equation}
reduces the summation over the ``azimouthal'' index $ \alpha$  

\begin{equation}
\sum_{\alpha=0}^{N}
Y_{\alpha}^N(\frac{a_{{\eta}'}}{a_B}{\eta})
Y_{\alpha}^N(\frac{a_B}{a_{{\eta}'}}{\eta}') = 
\sum_{\alpha=0}^{N}
Y_{\alpha}^N({\eta})Y_{\alpha}^N({\eta}') 
\end{equation}
since 

$$ \frac{a_{{\eta}'}}{a_B}{\eta}.\frac{a_B}{a_{{\eta}'}}{\eta}'={\eta}.{\eta}'$$

In effect, the expansion (14) for the boundary part of $ D_{c}^n(\eta,{\eta}')$ 
reduces to that in (9) with the stated range of summation over $ N$ between $ 0$ 
and $ N_0$. This reduction allows for manipulations of the boundary part of 
$ D_{c}^n(0)$ in transform space similar to those of $ D_{s}^n(0)$. It becomes 
evident, nevertheless, that whereas on $ S_n$ the massless propagator vanishes
because the $ N=0$ term - which in de Sitter space is associated with the 
infra-red divergence - cancels identically against the ultra-violet sum-total of 
the remaining infinite number of terms \cite{Drummond} the image propagation on 
$ S_n$ expressed by (14) entails only a finite number of terms in the relevant 
summation over $ N$. Consequently, the same exact cancellation is no longer possible 
on $ \it{C_n}$. The result, albeit finite to zero (bubble) and first order (tadpole)
in the renormalised self-coupling $ \lambda_R$, is of crucial importance to 
renormalisation as radiative effects at higher orders will generate ultra-violet 
divergences for the tadpole diagram. In turn, the presence of $ \partial C$ may 
significantly affect the renormalisation program of conformal scalar fields on a 
Riemannian manifold of constant curvature with the potential for generation of 
mass and conformally non-invariant counterterms to orders at which the latter 
have been shown to be absent on $ S_n$. In this respect, it is imperative to 
explore the potential contributions of all $ D_{c}^n(0)$-entailing diagrams to 
the effective action
 
\begin{figure}[h]
\centering\epsfig{figure=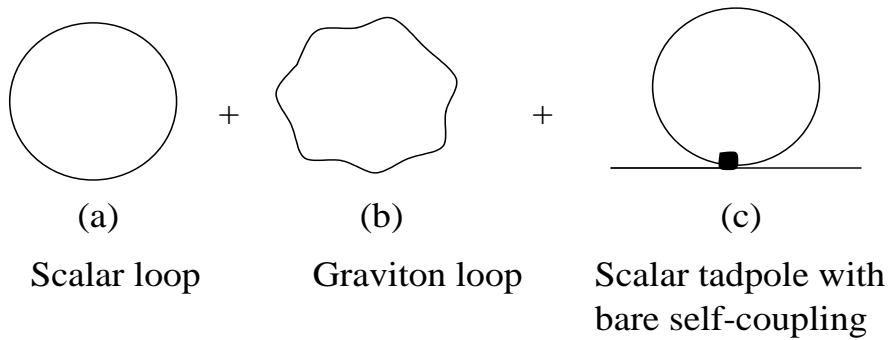, height = 45mm,width=120mm}
\caption{$D_{c}^n(0)$-related contributions to the effective action}
\end{figure}

The ``bubble'' diagrams in fig.(1a) and fig.(1b) not entailing any self-couplings 
are finite in the context of dimensional regularisation. However, in general power 
counting terms, they account diagramatically for the simultaneous one-loop 
contributions to volume and boundary effective Einstein-Hilbert action on any 
manifold with boundary. Such simultaneous contributions at one loop-level
in the absence of self-couplings have been assessed in the general case of 
non-minimal matter through heat kernel techniques \cite{Solodukhin}. The 
Einstein-Hilbert action is, in the case of minimal coupling of matter to gravity, 
the semi-classical gravitational 
component of the effective action for a scalar field coupled to the background 
geometry of $ \it{C_n}$ and at $ n=4$ it assumes the familiar form

\begin{equation}
S_{EH} = -\frac{1}{16\pi G}[\int_{C}{d\sigma}(R-2\Lambda)+ 
2\int_{{\partial}C}{d^3}x \sqrt{h}K]
\end{equation}
with $ d\sigma = a^nd\Omega_{n+1}$ being the element of surface area of the 
n-sphere embedded in $ n+1$ dimensions and with the three-dimensional boundary 
hypersurface on which the induced metric is $ h_{ij}$ being characterised by 
an extrinsic curvature $ K_{ij}=\frac{1}{2}(\nabla_{i}n_{j}+\nabla_{j}n_{i})$
the trace of which is $ K=h^{ij}K_{ij}$. The surface term in (21) has been the result 
of distinct approaches to field theoretical formulations on a manifold with boundary
\cite{York}. In a variational context it was introduced by Gibbons and Hawking \cite{GH}
in order to offset the variations of the normal derivatives of the metric on the
boundary stemming from the variation of the volume term of the classical action in
the context of a fixed metric on the boundary. 

The unique conformally invariant bare scalar action on a n-dimensional 
Riemannian manifold of constant curvature $ \it{a}$ is \cite{McKeon Tsoupros}
 
\begin{equation}
S[{\Phi}_0] = \int_C{d\sigma}[\frac{1}{2} \frac{1}{2a^2}
\Phi_0(L^2- \frac{1}{2}n(n-2) )\Phi_0 - \frac{\lambda_0}{\Gamma(p+1)}\Phi_0^p]
\end{equation}
provided that $ p = \frac{2n}{n-2}, n>2$. $ L_{\mu \nu}$ is the generator of 
rotations 

\begin{equation}
L_{\mu \nu} = \eta_{\mu}\frac{\partial}{\partial \eta_{\nu}} -
\eta_{\nu}\frac{\partial}{\partial \eta_{\mu}} 
\end{equation}
on the embedded sphere. In the physically relevant case of $ n=4$ the self-coupling
is that of $ \Phi^4$. $ \it{C_n}$ is characterised by spherical $(n-1)$-dimensional 
sections of constant Euclidean time $ \tau$. In addition to (22) the presence of a 
boundary on the Riemannian manifold of constant curvature necessitates at the 
classical level a term of the form $ \int_{\partial C} K{\Phi^2}$. In the same 
variational context which, in the case of minimal coupling, 
gives rise to the Gibbons-Hawking boundary term in (21) this 
term is necessary in order to eliminate the non-vanishing variations of the normal 
derivative on the boundary stemming from the non-minimal coupling of the conformal
scalar field to the background metric. It should be stressed, in addition, that this 
term which replaces the surface term in (21) 
is present in the classical action despite the Dirichlet condition of 
$ \Phi(\eta)=0$ on the boundary since the boundary integral, essentially taken over 
the product of two scalar fields, is predicated on the coincidence limit 
$ \eta \rightarrow \eta'$ which independently results in a divergence in both terms 
of the propagator in (7) \cite{Solodukhin}. This term is, in principle, also
expected in the effective action at higher loop-orders despite the Dirichlet condition
since for any diagram the coincidence limit of integration points is the potential 
source of divergences on the manifold's volume as well as boundary. 
The issue which naturally arises in such a context is whether the 
effective action, obtained by integrating out the interacting quantum scalar field
$ \Phi$ reproduces at higher loop-orders the semi-classical boundary term in question.       

The ``tadpole'' diagram in fig.(1c) formally involving the bare self-coupling 
$ \lambda_0$ as well as bare scalar field $ \Phi_0$ has the potential for 
contributions to both the volume scalar effective action as well as to its boundary 
counterpart of the form $ \int{K \Phi^2}$. Being a diagramatic representation of the 
first non-trivial correction to the two-point function it yields a contribution to 
the propagator of the form

\begin{equation} 
\frac{1}{2}(- \lambda_0)
\int_{C}{d^n}{\eta}D_{c}^{(n)}(\eta_1,{\eta})D_{c}^{(n)}(\eta,{\eta})D_{c}^{(n)}(\eta,{\eta}_2)
\end{equation}
On the grounds of the preceding analysis the loop integral expressing the proper 
version of the same diagram (no external propagators attached to the loop) is given 
by (15), (7) and (14) and by virtue of (20) and of the absence of the 
fundamental-related part it assumes the form 

\begin{equation} 
\int_{C}{d^n}{\eta}D_{c}^{(n)}(\eta,{\eta}) = -\sum_{N=0}^{N_0}\sum_{\alpha=0}^{N}
\frac{1}{\lambda_N}Y_{\alpha}^N({\eta})Y_{\alpha}^N({\eta})
\int_{C}{d^n}{\eta}
\end{equation}     
which, in the context of (18), (19) and (13), yields \cite{Drummond}

$$
\int_{C}{d^n}{\eta}D_{c}^{(n)}(\eta,{\eta}) = 
$$

\begin{equation} 
-\frac{2N+n-1}{4\pi^\frac{n}{2}}\Gamma(\frac{n-1}{2})C_N^\frac{n-1}{2}(1)
a^{2-n}\sum_{N=0}^{N_0}\frac{(2N+n-1)\Gamma(N+n-1)}{(N+\frac{n}{2})(N+\frac{n}{2}-1)\Gamma(N+1)}
\int_{C}{d^n}{\eta}
\end{equation}                
The sum over the angular momentum quantum number $ N$ associated with image 
propagation in (26) amounts to a finite multiplicative factor. In fact, for
sufficiently small geodesic distances $ a_B$ between the $ \partial C_n$ and the 
n-cap's pole only the first few terms in this sum are relevant.

The volume integral featured in (26) has a well defined value on the Euclidean 
embedded n-sphere $ S_n$ of radius $ a$, namely \cite{Bateman}

\begin{equation}
\int_{S}{d^n}{\eta} = a^n \frac{2\pi^{\frac{n+1}{2}}}{\Gamma(\frac{n+1}{2})}
\end{equation}
On $ \it{C_n}$, however, the final integration over the angle $ \theta_n$ 
extends from 0 to the boundary-defining value of $ \theta_n^{(0)} < \frac{\pi}{2}$.
Consequently, 

$$
\int_{C}{d^n}{\eta} = a^n 
\int_0^{2\pi}d\theta_1\int_0^{\pi}d\theta_2sin\theta_2\int_0^{\pi}d\theta_3sin^2\theta_3...
\int_0^{\theta_n^0}d\theta_nsin^{n-1}\theta_n
$$

\begin{equation}
= a^n \frac{2\pi^{\frac{n}{2}}}{\Gamma(\frac{n}{2})}\int_0^{\theta_n^0}d\theta_nsin^{n-1}\theta_n
\end{equation}
At the limit of space-time dimensionality $ n=4$ this is

\begin{equation}
\int_{C}{d^4}{\eta} = a^4 \frac{2\pi^{2}}{\Gamma(2)}[-\frac{1}{3}sin^2\theta_4^0cos\theta_4^0
- \frac{2}{3}cos\theta_4^0 + \frac{2}{3}]
\end{equation}
This is the volume integral of $ \it{C_4}$ the metric of which manifold is

\begin{equation}
ds^2 = a^2[d\theta_4^2 + sin^2\theta_4d\Omega_3^2]
\end{equation}
with $ d\Omega_3^2$ being the metric of three spheres characterised by an 
extrinsic curvature the trace of which is  

\begin{equation}
k = \frac{1}{3}\frac{1}{a}cot\theta_4
\end{equation}

The boundary hypersurface of $ \it{C_4}$ is the three sphere of radius 
$ asin\theta_4^0$ and of extrinsic curvature the trace of which is  

\begin{equation}
K = \frac{1}{3}\frac{1}{a}cot\theta_4^0 
\end{equation}
with

\begin{equation}
K' \equiv \frac{dK}{d\theta}_{| \partial C} = 
-\frac{1}{3}\frac{1}{a}\frac{1}{sin^2\theta_4^0}
\end{equation} 

Together (29), (32) and (33) reduce (26) at $ n=4$ to

\begin{equation} 
\int_{C}{d^4}{\eta}D_{c}^{(4)}(\eta,{\eta}) =  
\int_{\partial C}d^3\eta K - 6a\int_{\partial C}d^3\eta KK' - 
\frac{2}{3}\frac{1}{a^2}\pi^{-\frac{1}{2}}\Gamma(\frac{5}{2})\int_{S_4}d^4\eta 
\end{equation} 
allowing for a multiplicative factor in each term which involves the sum in 
(26) and whose significance will be discussed in what follows.

It becomes evident through (24) and (34), for that matter, that, although the 
``tadpole'' diagram entails no primitive divergences in its structure it 
contributes simultaneously, at any order in the perturbative expansion of 
$ \lambda_0$, to both the volume bare action for $ \Phi_0$ and to its 
boundary counterpart. Allowing for inessential multiplicative factors the 
form of these contributions is 

\begin{equation}
(-\lambda_0)\int_{\partial C}d^3\eta K\Phi^2 - 2a(- \lambda_0)\int_{\partial C}d^3\eta KK'\Phi^2 
-(- \lambda_0)\int_{C - \partial C}d^4\eta R\Phi_0^2 
\end{equation}  
where in the last term use has been made of the Euclidean de Sitter space relation 
\cite{McKeon Tsoupros} 

\begin{equation}
R = \frac{n(n-1)}{a^2}
\end{equation}
and the fact that the third term in (34) amounts in its entirety to the volume 
integral over $ {\it C_4}$ excluding its boundary. On account of the stipulated 
Dirichlet condition the radiative contributions in (35) feature the 
semi-classical scalar field $ \Phi$ on $ \partial C$ as opposed to the bare $ \Phi_0$.  

As stated, physical requirements enforce the first term in (35) in the clasical action 
and semi-classical level of the quantum effective action. For that matter, it is
also expected at higher orders. However, at the coincidence limit of 
$ \eta \rightarrow \eta'$ on the boundary it is $ a_{\eta'}=a_{B}$ and,
as a consequence, the boundary-related term in (17) of the loop integral in (15) which 
is finite and non-vanishing in the volume reduces to the same form as that of the 
fundamental-part related term in (16). As a result, the expansion in the aforementioned 
multiplicative factor stemming from (26) in the first two terms of (35) is no longer 
truncated at $ N_0$. The peculiar cancellation between the ultra-violet and the 
infra-red divergence of the massless ``tadpole'' in dimensional regularisation and the 
concomitant vanishing effect is again ``at work'' on the boundary forcing the first 
two terms in (35) to vanish. As a consequence, the ``tadpole's'' contribution to the 
bare scalar action on $ {\it C_4}$ and to any order in the perturbative expansion of 
the bare self-coupling reduces to only the third term in (35). This result is consistent 
at one-loop order with the absence of matter-related surface counterterms in the effective 
action on a general manifold with boundary \cite{Odintsov}.  

Allowing for the stated vanishing effect it is, nevertheless, worth noting that 
the dimensionally consistent form of the radiatively induced terms in (35) as well as
the general form of the volume integral in (29) which characterises all Feynman diagrams 
on $ \it{C_4}$ suggest in the context of (32) and (33) that radiative contributions to
the two-point function tend to reproduce the form of the classical boundary term in the 
quantum effective action already at low orders in perturbation. These radiative 
contributions at $ n \rightarrow 4$ stem exclusively from the bare $ \Phi_0$ and bare 
self-coupling $ \lambda_0$ which respectively admit an expansion in terms of 
$ \Phi$ and $ \lambda$ \cite{'t Hooft} as

\begin{equation}
\Phi_0 = Z^{\frac{1}{2}}\Phi~~~;~~~ 
Z = 1+\sum_{k=1}^{\infty}\sum_{i=1}^{\infty}\frac{c_{ki}\lambda^i}{(4-n)^k}
\end{equation}

\begin{equation}
\lambda_0 = \mu^{4-n}[\lambda + \sum_{\nu=1}^{\infty}\frac{a_{\nu}(\lambda)}{(4-n)^{\nu}}]
\end{equation}
and allow, in the presence of self-couplings, for the generalisation of the 
statement in \cite{Solodukhin} to the effect that both volume and 
boundary-related ultra-violet divergences are simultaneously removed. In the absence 
of self-couplings such a simultaneous process is realised through the renormalisation
of the gravitational coupling $ G$ in the volume sector of (21). In the present case, 
in addition to the same renormalisation of $ G$, any counterterm to the ``tadpole'' 
diagram generated by overlapping divergences through (37) and (38) at any fixed order 
in $ \lambda$ contributes, in principle, simultaneously to all three sectors in (35). 
Disregarding the vanishing effect of the boundary terms in the ``tadpole'' case such 
simultaneous contributions appear to be relevant to any diagramatic 
structure at any higher-loop order in the perturbative expansion of the two-point 
function. As a result of such perturbative redefinitions at higher orders the bare 
action on $ {\it C_n}$ can be seen to feature the tree terms in (35) in addition to the
bare action on $ S_4$ expressed by (22) at $ n=4$ 

$$
S[{\Phi}_0] = (-\lambda_0)\int_{\partial C}d^3\eta K\Phi^2 - 2a(- \lambda_0)\int_{\partial C}d^3\eta KK'\Phi^2 
 + $$
\begin{equation}
\int_C{d\sigma}[\frac{1}{2} \frac{1}{2a^2}\Phi_0(L^2- \frac{1}{2}n(n-2) )\Phi_0 + \kappa_0 R\Phi_0^2  
- \frac{\lambda_0}{4!}\Phi_0^4]
\end{equation} 
with $ \kappa_0$ being a bare non-minimal coupling which vanishes at the semi-classical 
level. The effective action will necessarily have the same form as the bare action if the 
bare quantities are replaced by their renormalised counterparts as a result of the summation
over vacuum diagrams plus counterterms to any specific order \cite{Coleman}.

Although the exact nature of vacuum contributions to the effective action necessitates
a detailed renormalisation program to any specific order a qualitative assessment can be 
elicited from (39). As announced, radiative effects at higher loop orders will 
necessarily generate the boundary term $ \int K\Phi^2$ although the latter is irrelevant 
to the diagramatic structure of the ``tadpole'' in fig.(1c).  The additional 
$ a\int KK'\Phi^2$ term in the effective action is a purely quantum effect. A consistent 
variational procedure of the classical Einstein-Hilbert action does not necessitate it 
\cite{Solodukhin}. The present calculation suggests at higher orders the simultaneous 
emergence of the stated two terms on the boundary $ \partial C_4$ and the volume 
contributions represented by the $ \int R\Phi_0^2$ term in (35) and (39). This term is 
the unique contribution of vacuum effects to order one in loop and bare self-coupling 
expansion and deserves attention in its own merit. Following an argument in \cite{Drummond} 
the quadratic expression  involving the spherical d' Alembertian $ L^2 - \frac{1}{2}n(n-2)$ 
in (39) relates to wave-function renormalisation effected by contributions to the two-point 
function represented by diagrams of, at least, two self-coupling vertices. These two 
vertices are rotationally related through the operator given by (23). The ``tadpole'' 
diagram featuring one vertex does not fall in that category and that is reflected in the 
absence of that operator in its volume contribution. The $ R\Phi_0^2$ term in (39) is, for 
that matter, distinct from the quadratic first term in the same expression. Since that 
quadratic term is the spherically formulated sum of the kinetic $ \phi_0 \partial^2 \phi_0$ 
and conformal $ \frac{1}{6}R\phi_0^2$ sector for a conformal scalar field in the case of a 
general manifold the presence of the third volume term in the effective action does not 
reproduce perturbatively the classical conformal coupling of the scalar field to the 
background geometry of $ {\it C_4}$. This situation contrasts sharply with the absence of 
conformally non-invariant counterterms for the same massless scalar theory on 
$ S_4$ \cite{Drummond}. Since this radiative effect arises from the ``tadpole'' diagram 
which entails no primitive divergences in dimensional regularisation the lowest order in 
perturbation at which the 
generation of this conformally non-invariant counterterm in the effective action is 
expected is order two in the bare self-coupling expansion. This expectation is predicated 
on the first non-trivial ultra-violet divergence stemming from the correction at that order 
to the four-point function for the same scalar theory defined on $ S_4$ \cite{Drummond}. 
Since that volume divergence is a cut-off scale effect arising from the coincidence of 
integration points in configuration space it is expected to be topology-independent and 
persist at least as a leading divergence in the volume of $ C_4$. Consequently, to order 
two in the expansion of $ \lambda_0$ the tadpole structure will represent a divergent 
counterterm to the two point function. Conformal invariance is broken as a result of 
boundary-related contributions to the effective action. The conformal anomaly for the one-loop
effective action has been evaluated through use of heat kernel techniques \cite{Moss}. The 
explicit calculation of the conformal anomaly in the context of the above considerations as 
well as the additional suggestive aspect of this calculation   
relating to the generation of the two boundary terms in the effective action and to the 
simultaneous renormalisation of boundary and volume ultra-violet divergences will be 
explicitly confirmed through higher-loop renormalisation.  

\bigskip

\section*{\bf{ACKNOWLEDGEMENTS}}

I am grateful to my father for financial support which made this project possible.

\end{document}